\documentclass[a4paper,10pt]{article}
\usepackage[utf8x]{inputenc}
\usepackage[margin=0.8in]{geometry}
\usepackage{amsmath}
\usepackage{relsize}
\usepackage{braket}
\usepackage{amsfonts}
\usepackage{mathtools}
\usepackage{graphicx}
\usepackage{float}
\usepackage{caption}
\usepackage{amssymb}
\usepackage{enumerate}
\usepackage{cancel}
\usepackage{indentfirst}
\usepackage{bm}
\usepackage{textcomp}
\usepackage{changepage}
\usepackage{commands}
\usepackage{multirow}
\usepackage{makecell}
\usepackage{booktabs}
\usepackage{chngcntr}
\usepackage{subfigure}
\usepackage{grffile}
\usepackage{achemso}
\usepackage{adjustbox,lipsum}
\usepackage{cleveref}
\usepackage[title]{appendix}

\counterwithin{subsubsection}{section}

\graphicspath{{/home/vbacic/my_linear_response/paper/}}
\DeclareMathSizes{10}{10}{6}{4}

\title{Analytical Approach to Phonon Calculations in the SCC-DFTB Framework}
\author{Vladimir Ba\v{c}i\'{c},$^1$ Thomas Heine$^{2,3,4}$ and Agnieszka Kuc$^2$\\
$^1$Department of Physics and Earth Sciences, Jacobs University Bremen,\\ Campus Ring 1, 28759 Bremen, Germany\\
$^2$Helmholtz-Zentrum Dresden-Rossendorf, Abteilung Ressourcen\"{o}kologie, Forschungsstelle Leipzig,\\ Permoserstr. 15, 04318 Leipzig, Germany\\
$^3$Theoretical Chemistry, TU Dresden, Mommsenstr. 13, 01062 Dresden, Germany\\
$^4$Department of Chemistry, Yonsei University,\\ 50 Yonsei-ro, Seodaemun-gu, Seoul 03722, Korea}

\date{}

\begin{document}

\maketitle

\begin{abstract}
Detailed derivation of the analytical, reciprocal-space approach of Hessian calculation 
within the self-consistent-charge density functional based tight-binding framework (SCC-DFTB) is presented.
This approach provides an accurate and efficient way for obtaining the SCC-DFTB Hessian of periodic systems.
Its superiority with respect to the traditional numerical force differentiation method is demonstrated
for doped graphene, graphene nanoribbons, boron-nitride nanotubes, bulk zinc-oxide and other systems.
\end{abstract}

\section{Introduction}
In the past two decades, density-functional based tight-binding method (DFTB) \cite{Porezag1995, Seifert1996, Elstner1998, Oliveira2009} has become a relatively popular tool for quantum mechanical simulations of 
large systems, otherwise computationally too demanding for the standard density-functional theory (DFT) or other ab-initio methods.
Introduced in the mid-90s as an approximation to DFT, DFTB has been subjected to ongoing extensions and improvements, 
the so-called self-consistent charge (SCC) DFTB \cite{Elstner1998} being perhaps the most important one.
In spite of its shortcomings, DFTB (along with its extensions) has shown to perform reasonably well for a variety of systems, 
sometimes even with an accuracy comparable to that of its first-principles counterpart, with only a fraction of the computational cost \cite{Gaus2013}.
Having this in mind, and motivated by the need of having an efficient and reasonably accurate way for studying the vibrational properties of large systems, Witek et al.\ \cite{Witek2004},
and subsequently Nishimoto and Irle \cite{Nishimoto2017}, developed an analytical method for obtaining the Hessian 
(i.e., matrix of geometrical second derivatives, needed in calculation of vibrational frequencies), within the SCC-DFTB framework.
Although considerably more efficient than the traditional numerical force differentiation method of Hessian calculation, the application of the approach from refs.\ \cite{Witek2004} and \cite{Nishimoto2017}
to periodic systems requires using supercells, which results in lower overall accuracy and higher computational costs.
\\
\indent
In this paper, we present an analytical and \textit{supercell-free} method for calculating the SCC-DFTB Hessian of periodic systems.
The underlying approach is based on the direct evaluation of the discrete Fourier transform of the Hessian (rather than the Hessian itself),
which can be achieved by taking the second derivative of the total energy with respect to collective and phase-modulated atomic displacements.
This \textit{reciprocal-space} method of phonon calculations is not new, in fact, it has been implemented in the plane-wave and muffin-tin based DFT codes long time ago, 
and is also known as the density-functional perturbation theory (DFPT) or linear-response (LR) theory \cite{Giannozzi1991, Savrasov1992}.
The objective of this work is to develop the SCC-DFTB analogue of such an approach to phonon calculations.
\\
\indent
This paper is organized as follows: Section \ref{general} provides a short review of the standard DFTB equations, along with the notation used throughout the rest of the paper.
In Section \ref{d_enrg}, we discuss the main idea behind the reciprocal-space approach to Hessian calculation.
In Section \ref{d2E_DFTB}, we present a derivation of the expression for evaluating the Fourier-transformed Hessian within SCC-DFTB.
For the sake of clarity, only the most important steps are shown here, while complete and extensive mathematical details can be found in the Supplementary Material.
Comparison of the reciprocal-space and the numerical force differentiation method of phonon calculation, along with the underlying discussion, is given in Section \ref{R&D},
while some concluding remarks are given in Section \ref{conclusion}.
\\
The theoretical formalism developed in this work has been implemented in a locally modified version of DFTB program in Amsterdam Modeling Suite\cite{AMS2018}, version 2018.
\newpage
\section{Theory}
\subsection{General DFTB Formalism}\label{general}
We begin by giving a brief overview of the SCC-DFTB framework, without going too deep into details. 
A more thorough discussion on this topic, as well as the derivations of the equations presented in this Section, can be found in the literature \cite{Koskinen2009, Oliveira2009}.
In the SCC-DFTB approximation, the Kohn-Sham (KS) energy functional for periodic systems can be written as:
\begin{equation}\label{DFTBeq}
  \mathcal{E} \left[ \{ \ckn \} \right]
    = \sum_{\vec{k}, n} f_{\vec{k}, n}^{} \, \ckndag \, \Hknull \, \ckn    \,+\,
	\frac{1}{2} \sum_{I,J} \sum_{\vec{R}} \gamma_{IJ}^{}(\vec{R}) \, \varDelta z_I^{} \varDelta z_J^{}  \,+\,
      \frac{1}{2} \sum_{I,J} \sum_{\vec{R}} V_{IJ}^{rpl}(\vec{R})
\end{equation}
In the first term, $\vec{k}$ is a vector in the Brillouin zone, $n$ is the band index, $\ckn$ is a column-vector of the orbital coefficients, 
$\Hknull$ is the parametrized Hamiltonian matrix (in the Bloch basis) and $f_{\vec{k}, n}$ is the electron occupation function.
In the last two terms, the sum runs over all atom pairs $(I,J)$ and all lattice vectors $\vec{R}$.
In the second term, the function $\gamma_{IJ}$ describes the Coulomb interaction of atomic charge fluctuations $\varDelta z$.
Finally, $V_{}^{rpl}$ is the so-called repulsion potential, parametrized as a short-ranged isotropic force-field.
The three terms in eq.\ \eqref{DFTBeq} are called band-structure ($\mathcal{E}_{BS}$), charge-fluctuation ($\mathcal{E}_{CF}$) and repulsion energy ($\mathcal{E}_{rpl}$), respectively. 
We note that $\vec{H}_{\vec{k}}^0$, $\gamma_{IJ}$ and $V_{}^{rpl}$ depend only on the pre-calculated parameter set and the geometry of the system, 
but not on the orbital coefficients. \\
\indent
The charge fluctuations in \eqref{DFTBeq} are most commonly calculated using the Mulliken population analysis \cite{Mulliken1955}.
According to it, the charge fluctuation on atom $I$ is given by:
\begin{equation}\label{z_mull}
  \varDelta z_I^{} = \frac{1}{2} \sum_{\vec{k},n} \fkn \, \sum_{a \in I} \sum_{b} 
                    \left( c_{\vec{k}, n}^{a *} \, c_{\vec{k}, n}^{b} S_{\vec{k}}^{ab} +  c_{\vec{k}, n}^{b *} \, c_{\vec{k}, n}^{a} S_{\vec{k}}^{ba}  
                    \right) - z_{I}^0
\end{equation}
where $z_I^0$ is the valence charge of the corresponding neutral atom, $S_{\vec{k}}^{ab}$ is the overlap matrix element between atomic basis functions $a$ and $b$ (also in the Bloch basis).
To avoid explicitly writing the double sum over the basis functions, it is convenient to introduce 
the atom projection matrix $\PI$ and the charge projection matrix $\vec{Z}_{I, \vec{k}}^{}$, defined by:
  \begin{align}
    \mathcal{P}_{\hspace{-0.20em}I}^{ab} & \equiv
    \begin{dcases}
       \delta_{a,b} \text{, if basis functions } a \text{ and } b \text{ belong to atom } I \\
       0 \text{ ~~, otherwise} 
    \end{dcases} \label{P_def}
  \\
    \vec{Z}_{I, \vec{k}}^{} &  \equiv \frac{1}{2} \left( \PI \Sk + \Sk \PI \right) \label{Z_def}
  \end{align}
Using \eqref{P_def} and \eqref{Z_def}, the expression for the Mulliken charge fluctuations \eqref{z_mull} can be compactly written as:
\begin{equation}\label{Mull2}
   \begin{split}
      \varDelta z_I^{} & = \frac{1}{2} \sum_{\vec{k},n} \fkn \, \ckndag \left( \PI \Sk + \PI \Sk \right) \ckn - z_{I}^0
                       \\
                       & = \sum_{\vec{k},n} \fkn \, \ckndag \, \vec{Z}_{I, \vec{k}}^{} \, \ckn - z_{I}^0
   \end{split}
\end{equation}
\indent
According to the variational principle, the SCC-DFTB ground state energy is obtained by minimizing \eqref{DFTBeq} with respect to orbital coefficients, 
under the orthonormalization constraints: $\vec{c}_{\vec{k},m}^{\dag} \, \Sk \, \vec{c}_{\vec{k}, n}^{} \hspace{-0.20em}= \delta_{m,n}^{}$. 
This leads to a system of generalized eigenvalue equations for $\ckn$:
\begin{equation}\label{secular}
   \Big[ \vec{H}_{\vec{k}}^{0} \,+\, \sum_I V_I \, \vec{Z}_{I, \vec{k}} \Big] \ckn = \ekn \Sk \ckn
\end{equation}
where the eigenvalues $\ekn$ are the single-particle band-structure energies, and
\begin{equation}\label{V_I}
  V_I \equiv \sum_J \sum_\vec{R} \gamma_{IJ}(\vec{R}) \varDelta z_J^{}
\end{equation}
is the electrostatic potential on atom $I$ due to charge fluctuations on all atoms.
The term in the square brackets of \eqref{secular} can be regarded as the total Hamiltonian matrix $\vec{H}_{\vec{k}}^{}$.
As $\vec{H}_{\vec{k}}^{}$ includes the term with the charge fluctuations, which depend on the orbital coefficients, 
equations \eqref{Mull2} and \eqref{secular} have to be solved self-consistently, just like the standard DFT KS equations.
The SCC-DFTB ground-state energy is then given by evaluating the expression \eqref{DFTBeq} with the self-consistent orbital coefficients and charge fluctuations.\\
\subsection{Energy derivatives and vibrational properties}\label{d_enrg}
In the Born-Oppenheimer approximation, the frequencies and modes of phonons with wavevector $\vec{q}$ are
given as the eigenvalues and eigenvectors of the so-called dynamical matrix $\textbf{D}_{\vec{q}}$, defined as:
\begin{equation}\label{dynmat}
  \text{D}_{\vec{q}}^{(B, \beta ; A, \alpha)} \equiv \frac{1} {\sqrt{M_B^{} M_A^{}}} \sum_{\vec{R}} e_{}^{i\vec{qR}} \Phi_{\vec{R}}^{(B,\beta; A, \alpha)}
\end{equation}
where capital and Greek indexes denote atoms and Cartesian displacements, respectively, 
$M$ is the atomic mass and:
\begin{equation}\label{hess_def}
   \Phi_{\vec{R}}^{(B,\beta; A, \alpha)} \equiv \frac{\partial^2 E}{\partial u_{B,\vec{R}}^{\beta} \, \partial u_{A,\vec{0}}^{\alpha}}
\end{equation}
where $u_{X, \vec{R}}^{\mu}$ denotes the $\mu$-th Cartesian component of the position of atom $X$ belonging to lattice point $\vec{R}$,
is known as the interatomic force constant matrix, Hessian matrix or simply Hessian.
Although $\vec{\Phi}_{\vec{R}}$ is formally defined for all points of the Bravais lattice $\{ \vec{R} \}$ of the system, 
in practice, it has non-negligible values only on some finite subset of $\{ \vec{R} \}$, i.e., on a supercell of the underlying system.
Since the dynamical matrix is just the discrete Fourier transform of the Hessian, weighted by the inverse square root of atomic mass products, 
calculating the Hessian poses the main challenge in the study of vibrational properties and related phenomena.
\\
\indent
The conceptually easiest approach to this problem consists of numerical evaluation of the first-order force derivatives with respect to atomic displacements.
Although simple, this technique of Hessian calculation can be quite slow and inefficient, since it requires doing a number of force calculations on a supercell on which the Hessian is non-negligible,
thereby typically resulting in much higher computational costs compared to calculations on the corresponding primitive unit cell.
Numerical instabilities associated with numerical evaluation of derivatives can also present a more severe issue in this case.
Even if an analytical expression for evaluating the force derivatives is available, the problem of using supercells is still present.
\\
\indent
An alternative approach to phonon calculation is based on evaluating the discrete Fourier transform of the Hessian directly, 
namely by using the following identity:
\\
\begin{equation}\label{d2qE}
   \widetilde{\Phi}_{\vec{q}}^{(B,\beta; A,\alpha)} = \tilde{\partial}_{\text{-} \vec{q}}^{B, \beta} \tilde{\partial}_{\vec{q}}^{A, \alpha} E
\end{equation}
where
\begin{equation}\label{dq_def}
   \tilde{\partial}_{\vec{q}}^{X, \mu}  \equiv \sum_{\vec{R}} e_{}^{i\vec{qR}} \frac{ \partial }{ \partial u_{X, \vec{R}}^{\mu} } 
\end{equation}
is the discrete Fourier-transform of the position derivative operator.
Since \eqref{d2qE} is valid for any $\vec{q}$-point of the reciprocal space, the entire phonon spectrum can be obtained without calculating $\boldsymbol{\Phi}_{\vec{R}}$ at all.
In practice, however, it is generally much more efficient to evaluate $\widetilde{\boldsymbol{\Phi}}_{\vec{q}}$ on a regular grid of the Brillouin zone (often referred to as the q-grid),
apply the inverse Fourier transformation to get $\boldsymbol{\Phi}_{\vec{R}}$, and then use \eqref{dynmat} to compute the dynamical matrix at arbitrary $\vec{q}$-point.
From the properties of the discrete Fourier transform, it follows that the density of the q-grid determines the size of the supercell on which $\boldsymbol{\Phi}_{\vec{R}}$ is defined \cite{Visvesvara2008}.
Whether there is an actual advantage to this (i.e., reciprocal-space) approach to Hessian calculation, depends on the way the total energy is calculated.
In other words, for this approach to be useful for a given energy calculation method, one must be able to efficiently evaluate the right-hand side (RHS) of \eqref{d2qE} within that particular method. 
In the following Section, we shall see that within SCC-DFTB, this can in fact be done \textit{analytically and without using supercells}, just like in plane-wave and mixed-basis DFT formalisms.
\\
\indent
Before moving on, we briefly focus on the Fourier-transformed atomic position derivative operator \eqref{dq_def}.
If $F_{\vec{R}}$ is any atomic-position-dependent quantity defined on all lattice points of a periodic system (e.g., orbital coefficients, charge fluctuations etc.), 
then acting with $\tilde{\partial}_{\vec{q}}$ on it results in a phase-modulated quantity:
\begin{equation}\label{d_F}
   \tilde{\partial}_{\vec{q}}^{A, \mu} F_{\vec{R}}^{} =    e_{}^{i\vec{qR}} \sum_{\vec{R'}} e_{}^{i\vec{qR'}} \frac{ \partial F_{\vec{R}}^{} } { \partial u_{A, \vec{R + R'}}^{\mu}} 
                                                    \equiv e_{}^{i\vec{qR}} F_{\vec{R}}^{\Aq}
\end{equation}
But no such modulation is present when acting on $F_{\vec{R}}$ with two Fourier-transformed derivative operators with opposite wave-vectors:
\begin{equation}\label{d2_F}
  \tilde{\partial}_{\text{-} \vec{q}}^{B, \nu} \tilde{\partial}_{\vec{q}}^{A, \mu} F_{\vec{R}}^{} = F_{\vec{R}}^{\BmqAq}
\end{equation}
On the far RHS of \eqref{d_F} and \eqref{d2_F}, the indexes of Cartesian displacements in the superscript have been dropped for clarity.
Unless specified otherwise, such notation shall be used from now on, i.e., 
any quantity with a superscript containing the atom index and wave-vector pair(s) shall be assumed to be 
the phase-independent part of the quantity obtained by applying the Fourier-transformed derivative operator(s).
\subsection{Second derivatives of SCC-DFTB energy}\label{d2E_DFTB}
In this Section, we derive the expression for evaluating the RHS of \eqref{d2qE}, for the case where $E$ is the SCC-DFTB total energy.
Following refs.\ \cite{Savrasov1992} and \cite{Gonze1995}, we do this by applying the variational principle to the second derivative of the DFTB KS energy functional \eqref{DFTBeq}:
\begin{equation}
   \tilde{\partial}_{\text{-}\vec{q}}^{B, \beta} \tilde{\partial}_{\vec{q}}^{A, \alpha} E = 
   \text{min} \Big[ \tilde{\partial}_{\text{-}\vec{q}}^{B, \beta} \tilde{\partial}_{\vec{q}}^{A, \alpha}\, \mathcal{E} \Big]
\end{equation}
This way, the expressions for the second derivatives of each component of $\mathcal{E}$ (band-structure, charge-fluctuation and repulsion) can be derived separately, so this is how we proceed. 
To keep the discussion as clear and as simple as possible, only systems with a finite band-gap and integer electron occupations 
(i.e., insulators and semiconductors at zero electron temperature) shall be considered here,
while the corresponding equations for the general case are given in the Supplementary Material.
\subsubsection{Second derivatives of $\mathcal{E}_{BS}$}
The second derivative of the band-structure part of $\mathcal{E}$ reads:
\begin{equation}\label{d2qE_BS}
  \begin{split}
   \dmqB \dqA \, \mathcal{E}_{\text{BS}}^{} = 
       \mathlarger{\sum}_{\kn} \, \fkn &
           \bigg[ \vec{c}_{\kn}^{\BmqAq \dag} \, \vec{H}_{\vec{k}}^{}           \, \ckn  \,+\, 
                  \ckndag                     \, \vec{H}_{\vec{k}}^{0 \, \BmqAq}   \ckn  \,+\,  
                  \ckndag                     \, \vec{H}_{\vec{k}}^{0}          \, \vec{c}_{\kn}^{\BmqAq} \,+
                 \\[-1\jot]
                 \phantom{sum_k} & \quad
                  \vec{c}_{\kn}^{\Bq \dag}   \, \vec{H}_{\vec{k}}^{0 \, \Aq }         \vec{c}_{\kn}^{}     \,+\,
                  \vec{c}_{\kn}^{\Bq \dag}   \, \vec{H}_{\vec{k+q}}^{0}            \, \vec{c}_{\kn}^{\Aq}  \,+\,
                  \vec{c}_{\kn}^{\dag}       \, \vec{H}_{\vec{k}}^{0 \, \Bq \dagger}  \vec{c}_{\kn}^{\Aq}  \,+
                 \\[1\jot]
                 \phantom{sum_k} & \quad
                  \vec{c}_{\kn}^{\Amq \dag}  \, \vec{H}_{\vec{k}}^{0 \, \Bmq }         \ckn                  \,+\,
                  \vec{c}_{\kn}^{\Amq \dag}  \, \vec{H}_{\vec{k-q}}^{0}             \, \vec{c}_{\kn}^{\Bmq}  \,+\,
                  \vec{c}_{\kn}^{\dag}       \, \vec{H}_{\vec{k}}^{0 \, \Amq \dagger}  \vec{c}_{\kn}^{\Bmq}
                 % \,+\,                    \Big( A \leftrightarrow B \quad \vec{q} \leftrightarrow \vec{-q}    \Big)
           \bigg]
  \end{split}
\end{equation}
The matrix elements of $\vec{H}_{\vec{k}}^{0 \, \Aq}$ and $\vec{H}_{\vec{k}}^{0 \, \BmqAq}$ are given by:
\begin{subequations}\label{dH0}
   \begin{gather}
      \left[ \vec{H}_{\vec{k}}^{0 \, \Aq} \right]_{ab} 
         = 
         \sum_{\vec{R}} \nabla H_{ab}^{0} (\vec{R} + \vec{u}_{ab}^{})
           e_{}^{i\vec{kR}} \Big( -\delta_{a}^{A} \,+\, e_{}^{i\vec{qR}} \delta_{b}^{A} \Big)
      \\
      \left[ \vec{H}_{\vec{k}}^{0 \, \BmqAq} \right]_{ab} 
         =
         \sum_{\vec{R}} \nabla^2 H_{ab}^{0} (\vec{R} + \vec{u}_{ab}^{}) e_{}^{i\vec{kR}}
           \Big( \delta_{A,B}^{} \left( \delta_{a}^{A} + \delta_{b}^{A} \right)   
                -e_{}^{-i\vec{qR}} \delta_{a}^{A} \delta_{b}^{B} \,-\, 
                 e_{}^{ i\vec{qR}} \delta_{b}^{A} \delta_{a}^{B}
           \Big)
   \end{gather}
\end{subequations}
where $\vec{u}_{ab}^{}$ is the bond vector between atoms to which basis functions $a$ and $b$ belong,
and $\delta_{x}^{X} {=} 1$ if basis function $x$ belongs to atom $X$, zero otherwise.
A completely equivalent expression holds for the matrix elements of $\SkAq$ and $\SkBmqAq$ as well. 
It is also worth mentioning that the RHS in \eqref{dH0} can in fact be evaluated analytically (see appendix \ref{App A}),
which contributes to the overall efficiency and accuracy of the reciprocal-space approach.
%which is beneficial in practical implementations.
%
\subsubsection{Second derivatives of $\mathcal{E}_{CF}$}
The second derivative of the charge-fluctuation part of $\mathcal{E}$ reads:
\begin{equation}\label{d2E_CF}
  \begin{split}
   \dmqB \dqA \mathcal{E}_{\text{CF}}^{} = \mathlarger{\sum}_{I,J}
                       &\Big[ \varDelta z_{I}^{\Aq}  \gamma_{IJ}^{\Bmq} \varDelta z_{J}^{}  +
                               \varDelta z_{I}^{\Bmq} \gamma_{IJ}^{\Aq}  \varDelta z_{J}^{}    \\[-3\jot]
                       & \; \; +  \,  
                               \varDelta z_{I}^{\Bmq} \, \widetilde{\gamma}_{IJ}^{} (\vec{q}) \, \varDelta z_{J}^{\Aq}    \,+\,
                               \varDelta z_{I}^{}     \, \widetilde{\gamma}_{IJ}^{} (\vec{0}) \, \varDelta z_{J}^{\BmqAq} 
                        \Big]
                      \, + \,
                       \dmqB \dqA E_{CF}^{} [\{ \varDelta z \}]
   \end{split}
\end{equation}
Here, $\widetilde{\gamma}_{IJ}^{}(\vec{q})$ is the phase-modulated lattice sum of $\gamma_{IJ}^{} (\vec{R})$:
\begin{equation}\label{gamma_lattsum}
  \widetilde{\gamma}_{IJ}^{} (\vec{q}) \equiv \sum_{\vec{R}} \gamma_{IJ}^{} (\vec{R}) e_{}^{i\vec{qR}}
\end{equation}
${\gamma}_{IJ}^{\Aq}$ is defined as:
\begin{equation}\label{d_gamma_lattsum}
   {\gamma}_{IJ}^{\Aq} \equiv \sum_{\vec{R}} \Big(  -    \delta_{J,A}^{} \nabla \gamma_{IA}^{} (\vec{R}) e_{}^{i\vec{qR}}
                                                   \,+\, \delta_{I,A}^{} \nabla \gamma_{AJ}^{} (\vec{R})
                                             \Big)
\end{equation}
and:
\begin{equation}\label{d2_gamma_E}
   \dmqB \dqA E_{CF}^{} [\{ \varDelta z \}] = \varDelta z_A^{} \sum_{\vec{R}}
      \Big( - \varDelta z_{B}^{}                        \nabla_{}^2 \gamma_{BA}^{}(\vec{R}) e_{}^{i \vec{qR}}
            + \delta_{A,B}^{} \sum_I \varDelta z_{I}^{} \nabla_{}^2 \gamma_{AI}^{}(\vec{R})
      \Big)
\end{equation}
is the second derivative of the charge fluctuation interaction with respect to atom positions only (i.e., keeping the charge fluctuations constant). 
The problem in \eqref{gamma_lattsum}-\eqref{d2_gamma_E} and \eqref{V_I} is that $\gamma$ is a long-ranged function (it decays as slowly as the Coulomb potential \cite{Gaus2013}), 
which makes the underlying lattice summations only conditionally convergent.
To overcome this issue, we use the well-known Ewald summation technique \cite{Ewald1921}, more details on this can be found in Appendix \ref{App B}.
\\
phase-independent part of the charge-fluctuation first derivative can be written 
in terms of orbital coefficients and their first derivatives as:
\begin{equation}\label{d_z}
  \varDelta z_{I}^{\Aq} =  \frac {1}{2} \sum_{\kn} \fkn 
                         \Big[  \, %\frac{1}{2}
                           \ckndag     \big( \PI \Skpq + \Sk   \PI \big) \vec{c}_{\kn}^{\Aq}  \,+\, 
                           \cknAmqdag  \big( \PI \Sk   + \Skmq \PI \big) \ckn
                         \Big] + \overline{\varDelta z}_I^{\, \Aq}
\end{equation}
with the last term defined as:
\begin{equation}
   \overline{\varDelta z}_I^{\, \Aq} \equiv \frac{1}{2} \, \sum_{\kn} \fkn \,  \ckndag \Big( \PI \SkAq +\, \vec{S}_{\vec{k}}^{\Amq \dagger} \PI \Big) \, \ckn
\end{equation}
Finally, the second derivative of the charge-fluctuations is given by:
\begin{equation}\label{d2_z}
  \begin{split}
  \varDelta z_{I}^{\BmqAq} = \frac{1}{2} \sum_{\kn} \fkn 
     \Big[ &
        \vec{c}_{\kn}^{\BmqAq \dagger} \, \vec{Z}_{I, \vec{k}}^{}             \, \vec{c}_{\kn}^{}       \,+\,
        \vec{c}_{\kn}^{\dagger       } \, \vec{Z}_{I, \vec{k}}^{}             \, \vec{c}_{\kn}^{\BmqAq} \,+\,
        \vec{c}_{\kn}^{\dagger       } \, \vec{Z}_{I, \vec{k}}^{\BmqAq}       \, \vec{c}_{\kn}^{}
        \\[-1pt] & \, +
        \vec{c}_{\kn}^{\Bq \dagger}    \, \vec{Z}_{I, \vec{k}}^{\Aq}          \, \vec{c}_{\kn}^{}     \,+\,
        \vec{c}_{\kn}^{\Bq \dagger}    \, \vec{Z}_{I, \vec{k+q}}^{}           \, \vec{c}_{\kn}^{\Aq}  \,+\,
        \vec{c}_{\kn}^{\dagger    }    \, \vec{Z}_{I, \vec{k}}^{\Bq \dagger}  \, \vec{c}_{\kn}^{\Aq}
        \\[5pt] & \, +
        \vec{c}_{\kn}^{\Amq \dagger}   \, \vec{Z}_{I, \vec{k}}^{\Bmq}         \, \vec{c}_{\kn}^{}     \,+\,
        \vec{c}_{\kn}^{\Amq \dagger}   \, \vec{Z}_{I, \vec{k-q}}^{}           \, \vec{c}_{\kn}^{\Bmq} \,+\,
        \vec{c}_{\kn}^{\dagger    }    \, \vec{Z}_{I, \vec{k}}^{\Amq \dagger} \, \vec{c}_{\kn}^{\Bmq}
     \Big]
   \end{split}
\end{equation}
where $\vec{Z}_{I, \vec{k}}^{\Aq}$ and $\vec{Z}_{I, \vec{k}}^{\BmqAq}$ are given exactly as in \eqref{Z_def}, with $\Sk$ replaced by $\SkAq$ and $\vec{S}_{\vec{k}}^{\BmqAq}$, respectively.
\subsubsection{Second derivatives of $\mathcal{E}_{rpl}$}
The second derivative of the repulsion energy is given by:
\begin{equation}\label{d2E_rpl}
  \dmqB \dqA E_{rpl}^{} = \sum_{\vec{R}} \Big( - \nabla_{}^2 V_{BA}^{rpl} (\vec{R}) e_{}^{i\vec{qR}} 
                                               + \delta_{A,B}^{} \sum_{I} \nabla_{}^2 V_{AI}^{rpl} (\vec{R})
                                         \Big)
\end{equation}
In contrast to $\gamma$, the DFTB repulsion potential is generally a short-ranged function which decays rapidly with distance, 
so evaluating the lattice on the RHS of \eqref{d2E_rpl} is straightforward.
Just like the repulsion energy, the repulsion contribution to the Hessian can be obtained independently from the band-structure and charge-fluctuation contributions.
\subsubsection*{Final expressions}
So far, we have derived expressions for the second derivatives of all components of $\mathcal{E}$. % the DFTB energy functional.
To proceed further, the second derivatives of the orthonormalization constraints must be taken into account as well:
\begin{equation}\label{d2q_orth}
  \begin{split}
     &      \vec{c}_{\kn}^{\BmqAq \dag} \,\Sk\, \ckn  \,+\, \ckndag \vec{S}_{\vec{k}}^{\BmqAq} \ckn   \,+\, \ckndag \Sk \, \vec{c}_{\kn}^{\BmqAq}   \\[3pt]
     & \quad + \,  \left( \cknBqdag \SkAq \, \ckn   \,+\, 
                          \cknBqdag \Skpq \, \cknAq \,+\, 
                          \ckndag   \vec{S}_{\vec{k}}^{\Bq \dagger} \, \cknAq    
                   \right) 
             + \Big( A \leftrightarrow B \quad \vec{q} \leftrightarrow \vec{-q} \Big) = 0
  \end{split}
\end{equation}
The left-hand side (LHS) of this equation is fully analogous to the term in the square brackets of \eqref{d2qE_BS}, with the Hamiltonian replaced by the overlap matrix. \\
Now, adding \eqref{d2qE_BS}, \eqref{d2E_CF} and \eqref{d2E_rpl}, while making use of \eqref{gamma_lattsum}-\eqref{d2_z} and \eqref{d2q_orth}, 
the second derivative of the SCC-DFTB energy functional becomes: 
\begin{equation}\label{d2E_dftb_var}
   \begin{split}
      \dmqB \dqA \mathcal{E} = 
        & \sum_{\kn} \fkn \bigg[     \cknBqdag \Big( \vec{H}_{\vec{k}}^{0 \Aq } + \sum_I V_I^{} \vec{Z}_{I, \vec{k}}^{\Aq } - \ekn \SkAq  \Big) \ckn  \\
        &        \hspace{1.5cm}  +\, \ckndag   \Big( \vec{H}_{\vec{k}}^{0 \Amq} + \sum_I V_I^{} \vec{Z}_{I, \vec{k}}^{\Amq} - \ekn \SkAmq \Big) \vec{c}_{\kn}^{\Bmq}  \\
        &        \hspace{1.5cm}  +\, \cknBqdag \Big( \vec{H}_{\vec{k+q}}^{} - \ekn \Skpq \Big) \cknAq
                        \bigg]
                        \,+\, \bigg[ A \leftrightarrow B \quad \vec{q} \leftrightarrow -\vec{q} \bigg]
         \\[5pt]
         & + \sum_{I,J} \bigg(  \gamma_{IJ}^{\Bmq}       \varDelta z_{I}^{\Aq}  \varDelta z_{J}^{}      \,+\, 
                                \gamma_{IJ}^{\Aq}        \varDelta z_{I}^{\Bmq} \varDelta z_{J}^{}      \,+\,
                     \widetilde{\gamma}_{IJ}^{}(\vec{q}) \varDelta z_{I}^{\Bmq} \varDelta z_{J}^{\Aq}   
                      \bigg)
         \\[5pt]
         & \,+\, \dmqB \dqA E_{BS}^{} [\{ \ckn \}] \,+\, \dmqB \dqA E_{CF}^{} [\{ \varDelta z \} ] \,+\, \dmqB \dqA E_{rpl}^{} \qquad
   \end{split}
\end{equation}
where:
\begin{equation}
   \dmqB \dqA E_{BS} [\{ \ckn \}] \equiv \sum_{\kn} \fkn \ckndag \Big( \vec{H}_{\vec{k}}^{0 \BmqAq } + \sum_I V_I^{} \vec{Z}_{I, \vec{k}}^{\BmqAq } - \ekn \vec{S}_{\vec{k}}^{\BmqAq}  \Big) \ckn
\end{equation}
just like $\dmqB \dqA E_{CF} [\{ \varDelta z \}]$, depends only on the orbital coefficients (but not on their derivatives!) and the geometry of the system.
\\
We see that the expression for $\dmqB \dqA \mathcal{E}$, as given by \eqref{d2E_dftb_var}, contains no second derivatives of either the orbital coefficients or the charge fluctuations.
Furthermore, it is variational with respect to the first derivatives of the orbital coefficients and, 
provided that the orbital coefficients on its RHS minimize $\mathcal{E}$, its minimum corresponds to the true value of the second derivative of the SCC-DFTB energy \cite{Gonze1995}.
\\
Varying \eqref{d2E_dftb_var} results in the following equation for $\vec{c}_{\kn}^{(A, \pm \vec{q})}$:
\begin{equation}\label{stern}
  -\Big( \vec{H}_{\vec{k \pm q}}^{} -\ekn \vec{S}_{\vec{k \pm q}}^{} \Big) \vec{c}_{\kn}^{\Apmq} = \Big( \vec{H}_{\vec{k}}^{\Apmq} - \ekn \vec{S}_{\vec{k}}^{\Apmq} \Big) \ckn
\end{equation}
and an equivalent one for $\vec{c}_{\kn}^{(B, \pm \vec{q})}$.
$\vec{H}_{\vec{k}}^{(A, \pm \vec{q})}$ is the matrix of the Hamiltonian total derivative, given by:
\begin{equation}
  \vec{H}_{\vec{k}}^{\Apmq} \equiv \vec{H}_{\vec{k}}^{0 \, \Apmq } + \sum_I \Big[ V_I^{} \vec{Z}_{I, \vec{k}}^{\Apmq} +
                                                                                  V_I^{\Apmq} \big( \PI \Sk + \vec{S}_{\vec{k \pm q}} \PI \big)
                                                                            \Big]
\end{equation}
with:
\begin{equation}
   V_I^{\Apmq} \equiv \sum_J \left( \gamma_{IJ}^{\Apmq} \varDelta z_J^{} + \widetilde{\gamma}_{IJ}^{} (\vec{\pm q}) \varDelta z_J^{\Apmq} \right)
\end{equation}
being the total derivative of the electrostatic potential.
In the literature, \eqref{stern} is also known as the Sternheimer equation \cite{Sternheimer1956}.
\\
\indent
So the problem of calculating $\dmqB \dqA E$ effectively reduces to the problem of determining the orbital coefficient derivatives.
These can be expressed as linear combinations of the orbital coefficients:
\begin{equation}\label{U_def}
   \vec{c}_{\kn}^{(X, \pm \vec{q})} = \sum_m U_{\vec{k} \, (n,m)}^{(X, \pm \vec{q})} \, \vec{c}_{\vec{k \pm q}, m}^{}
\end{equation}
where $\vec{U}_{\vec{k}}^{(X, \pm \vec{q})}$ is a square matrix (with the number of rows and columns equal to the number of atomic basis functions) to be determined.
From \eqref{stern}, it immediately follows that the entries of $\vec{U}_{\vec{k}}^{(X, \pm \vec{q})}$ matrix,
which refer to the non-degenerate pairs of states at $\vec{k}$ and $\vec{k \pm q}$ points, are given by:
\begin{equation}\label{U1}
  U_{\vec{k} \, (n,m)}^{(X, \pm \vec{q})} = \frac{ \vec{c}_{\vec{k+q}, m}^{\dag} \Big( \HkAq \,-\, \ekn \SkAq \Big) \,\ckn } 
                                                 { \ekn - \varepsilon_{\vec{k \pm q}, m}^{}                                }
\end{equation}
while for determining all other entries of $\vec{U}_{\vec{k}}^{(X, \pm \vec{q})}$, the following relation can be used (see S1):
\begin{equation}\label{d_orth1}
   U_{\vec{k \pm q} \, (m,n)}^{(X, \mp \vec{q}) *} \,+\, U_{\vec{k} \, (n,m)}^{(X, \pm \vec{q})} \,+\,\vec{c}_{\vec{k+q}, m}^{\dag} \, \SkAq \ckn = 0
\end{equation}
Inserting \eqref{U_def} to \eqref{d_z} and making use of \eqref{U1} and \eqref{d_orth1},
the following expression for charge-fluctuation derivatives is obtained:
%
%\begin{adjustwidth}{-0.5cm}{-0.5cm}
\begin{equation}\label{dz}
    \varDelta z_{I}^{\Aq} = \sum_{\vec{k}}  \sum_{n \in \mathcal{V}} 
      \left( \sum_{m \in \mathcal{C}} \frac{ \fkn M_{\vec{k} (m,n)}^{\Aq} } {\ekn - \ekpqm}  
           - 
             \frac{1}{2}
             \sum_{m \in \mathcal{V}} \fkpqm O_{\vec{k}(m,n)}^{\Aq} 
      \right) 
      \ckndag \big( \PI \Skpq +\, \Sk \PI \big) \ckpqm  \,+\, \overline{\varDelta z}_I^{\, \Aq} 
\end{equation}
%\end{adjustwidth}
%
Here, $\mathcal{V}$ and $\mathcal{C}$ refer to the sets of valence and conduction (i.e., occupied and empty) states, respectively,
while $M_{\vec{k} (m,n)}^{\Aq}$ is the (generalized) electron-phonon matrix element and $O_{\vec{k} (m,n)}^{\Aq}$ is the overlap derivative matrix element:
 \begin{equation}\label{M_def}
     M_{\vec{k} (m,n)}^{\Aq}  \equiv \vec{c}_{\vec{k+q}, m}^{\dag} \Big( \HkAq \,-\, \ekn \SkAq \Big) \,\ckn  \hspace{1in}
     O_{\vec{k} (m,n)}^{\Aq}  \equiv \vec{c}_{\vec{k+q}, m}^{\dag} \, \SkAq \ckn
 \end{equation}
It is easy to see that electron-phonon matrix elements and charge-fluctuation derivatives depend on each other, 
so they must be calculated self-consistently, much like the charge-fluctuations and the orbital coefficients.
At last, combining \eqref{d2E_dftb_var} with \eqref{U_def}-\eqref{d_orth1}, we arrive at the expression for the second derivatives of the SCC-DFTB energy:
%
%\begin{adjustwidth}{-0.5cm}{-0.5cm}
\begin{equation}\label{Phi_dftb}
   \begin{split}
      \dmqB \dqA E_{} = & \sum_{\vec{k}} \sum_{n \in \mathcal{V}} 
          \bigg[ 2\sum_{m \in \mathcal{C}} \fkn       \frac{ M_{\vec{k}(m,n)}^{\Aq} M_{\vec{k}(m,n)}^{\Bq *} } { \ekn - \ekpqm }
                 -\sum_{m \in \mathcal{V}} \fkpqm     \left( M_{\vec{k}(m,n)}^{\Aq} O_{\vec{k}(m,n)}^{\Bq *}  \, + \,  M_{\vec{k} (m,n)}^{\Bq*} O_{\vec{k}(m,n)}^{\Aq} \right) 
          \bigg]
          \\
          %
          %\;\; & + \fkn \ckndag \Big( \vec{H}_{\vec{k}}^{0 \, \BmqAq} + \sum_I V_I^{} \vec{Z}_{I, \vec{k}}^{\BmqAq} - \ekn \vec{S}_{\vec{k}}^{\BmqAq} \Big) \ckn
          %\bigg]
          \\[-5pt]
          & - \sum_{I,J}  \widetilde{\gamma}_{IJ}^{} (\vec{q}) \varDelta z_{I}^{\Bq *} \varDelta z_{J}^{\Aq}
            + \sum_I \Big[ \overline{\varDelta z}_I^{\, \Bq *} \, V_I^{\Aq}  \,+\, \overline{\varDelta z}_I^{\, \Aq *} \, V_I^{\Bq} \Big]
          \\
          & + \, \dmqB \dqA E_{BS}^{} [\{ \ckn \}] \,+\, \dmqB \dqA E_{CF}^{} [\{ \varDelta z \}] \,+\, \dmqB \dqA E_{rpl}^{}
   \end{split}
\end{equation}
%\end{adjustwidth}
%
\newpage
This expression, along with \eqref{dz}, is the main result of this paper.
Since all of the derivatives appearing in this Section can be evaluated analytically, the entire SCC-DFTB reciprocal-space approach to Hessian calculation can be considered analytical.
Although all expressions here are derived for periodic systems, they are also valid for non-periodic systems as well. 
This can be seen by taking the limit of infinitely large unit cells, thus restricting all real-space summations to a single unit cell 
and all $\vec{k}$ and $\vec{q}$-points to the $\Gamma$-point. 
In that case, reciprocal-space summations can be omitted, Fourier-transformed derivatives reduce to ordinary derivatives and
our entire formulation becomes equivalent to the one developed by Witek et al.\ \cite{Witek2004}.
In closing of this section, we once again point out that both \eqref{Phi_dftb} and \eqref{dz} are only valid for systems with a finite band gap and for vanishing electron temperature,
while the corresponding expressions for the case of arbitrary temperature can be found in the Supplementary Material ((S2.9) and (S2.11)). 
\newpage
\section{Test Calculations}
To test the performance of the reciprocal-space (analytical) approach to Hessian calculation and compare it to the traditional numerical force differentiation method, 
we used both approaches to compute the Hessians for a variety of systems of all dimensions.
In order to check how much the numerical and analytical results differ, 
we calculated the so-called root mean squared relative percentage difference \cite{Shcherbakov2013} of the Hessians resulting from the two approaches:
\begin{equation}\label{delta_phi}
  \Delta_{\mathcal{A, N}} \equiv \sqrt{ \frac{1}{(3N_{at})^2 N} \sum_{\vec{R}} \sum_{I,J} \left( \frac{\Phi_{\mathcal{A}, \, \vec{R}}^{(I;J)}} {\Phi_{\mathcal{N}, \, \vec{R}}^{(I;J)}} - 1 \right)^2 } \, \cdotp 100 \%
\end{equation}
where $\boldsymbol{\Phi}_{\mathcal{A}}$ and $\boldsymbol{\Phi}_{\mathcal{N}}$ are analytically and numerically obtained Hessians, respectively, 
$N_{at}$ is the number of atoms in the system and $N$ is the number neighbouring unit cells within the supercell on which the Hessians are defined.
In the analytical approach, this supercell is effectively determined by the q-grid used in the underlying calculation (as already mentioned in Section \ref{d_enrg}),
hence \eqref{delta_phi} (and comparing Hessians in general) only makes sense if the q-grid parameters used in the calculation of $\boldsymbol{\Phi}_{\mathcal{A}}$ 
are equal to the supercell parameters in the calculation of $\boldsymbol{\Phi}_{\mathcal{N}}$.
To make the comparison of both methods as consistent as possible, the k-grid parameters used in the reciprocal-space integration were made inversely proportional to the (super)cell size.
For example, if the Brillouin zone of some system was sampled with a $12 {\times} 12 {\times} 6$ k-grid, then a $6 {\times} 6 {\times} 3$ k-grid was used 
for doing calculations on a $2 {\times} 2 {\times} 2$ supercell, a $4 {\times} 4 {\times} 3$ k-grid for a $3 {\times} 3 {\times} 2$ supercell and so on.
\subsubsection*{Results and Discussion}\label{R&D}
Table \ref{RMSPD} shows examples of $\Delta_{\mathcal{A,N}}$ for cubic boron-nitride (zinc-blende phase), where the numerical Hessians were calculated with a different number of steps and step sizes.
We see that $\Delta_{\mathcal{A,N}}$ always drops when the step size is decreased,
whereas such a clear trend is not present when increasing the number of displacement steps.
The latter behavior can be attributed to the anharmonic effects, which are not captured by the analytical approach at this level of theory, 
but can always appear in the numerical approach for sufficiently large step sizes. 
In any case, it is clear that in the limit of small displacement step sizes, the numerically obtained Hessians converge to the analytically obtained one.
Similar behavior of $\Delta_{\mathcal{A,N}}$ is obtained for all other systems considered here, which confirms the accuracy of the analytic approach.
\begin{table}[h]
   \centering
   \caption{Root mean square relative percentage difference between analytically and numerically obtained Hessians
            ($\Delta_{\mathcal{A,N}}$, see eq.\ \eqref{delta_phi}) for zinc-blende BN \label{RMSPD}.
            Number of steps and step sizes refer to the parameters used in the numerical calculations.}
   \renewcommand{\arraystretch}{1.2} 
   \begin{tabular}{>{\centering\arraybackslash}c |c c c c }
     \hline
     \multirowcell{2}{total \# \\ of steps}  & \multicolumn{4}{c}{step size (Bohr)} \\ \cline{2-5}
                                             & 0.0150  & 0.0100  & 0.0050  & 0.0025 \\
     \hline
     \hline
     2                                       & 2.466   & 1.157   & 0.470   & 0.151  \\
     %\hline
     4                                       & 1.705   & 1.467   & 0.242   & 0.045  \\
     %\hline
     \multirowcell{1}{6}                     & 2.005   & 1.498   & 0.240   & 0.024  \\
     %\hline
     8                                       & 2.123   & 1.473   & 0.276   & 0.018  \\
     \hline
   \end{tabular}
\end{table}
\\
\indent
Hessian calculation timings for selected systems and different supercell (q-grid) parameters are given in table \ref{timings}.
For completeness, DFTB model without charge self-consistency (also known as DFTB0) was also included in the consideration.
We see that only for small systems and DFTB0 are the numerical and the analytical approach comparable in efficiency, 
while in all other cases the latter approach is much faster. 
The numerical to analytical timing ratios increase with the number of atoms in the system, as well as the supercell (q-grid) size, and are always greater in the SCC-DFTB case. 
This implies that the corresponding ratios would likely be even larger for a more sophisticated DFTB framework (such as DFTB3 \cite{Gaus2013}).
\\
\indent
When considering the overall efficiency of any quantum-chemical computational method, it is also important to take the aspects of parallelizability and symmetry into account. 
Since our implementation of the analytical method is currently serial and cannot make use of symmetry, all calculations (both numerical and analytical) were carried out on a single CPU core,
while disregarding all possible symmetries of the investigated systems. 
But even if the same calculations had been performed using a parallel implementation that does support symmetry, we expect that similar timing ratios would have been obtained, 
as both approaches are in principle parallelizable and can exploit symmetry to the same extent.
In the numerical case, for example, each force constant can be evaluated independently, while the number of independent force constants is determined by the symmetry of the system.
Likewise, in the analytical case each element of the $\widetilde{\boldsymbol{\Phi}}_{\vec{q}}$ matrix can be evaluated independently, 
while the number of independent q-grid vectors is also determined by the symmetry of the system \cite{Maradudin1968}.
\\
Finally, fig.\ \ref{fig:phdisp} shows examples of phonon spectra and phonon density of states of some systems from table \ref{timings},
calculated with SCC-DFTB.
\\
%\adjustbox{max width=\textwidth}{%
\begin{table}[h]
  \renewcommand{\arraystretch}{1.1}
  \caption{Hessian calculation timings (in seconds) for the analytical ($t_{\mathcal{A}}$) and numerical ($t_{\mathcal{N}}$) approaches. 
           $N_{at}$ is the number of atoms in the primitive unit cell of the given system and references refer to the parameter set used in the calculations. \label{timings}}
  \resizebox{\textwidth}{!}{%
  \begin{tabular}{l l r l r r r c r r r }
  \toprule
  \multirow{2}{*}{system}  & \multirowcell{2}{Ref.} & \multirowcell{2}{$N_{at}$ } & \multirowcell{2}{supercell \\ (q-grid)} & \multicolumn{3}{c}{SCC-DFTB} & &\multicolumn{3}{c}{DFTB0} \\ \cline{5-7} \cline{9-11}
                           &                        &                             &                                         &  $t_{\mathcal{N}}$ & $t_{\mathcal{A}}$ & $t_{\mathcal{N}} / t_{\mathcal{A}}$  &  
				                                                                                            &  $t_{\mathcal{N}}$ & $t_{\mathcal{A}}$ & $t_{\mathcal{N}} / t_{\mathcal{A}}$   \\
  \hline
  \hline
  \addlinespace
  \multirow{1}{*}{fullerene}                       & \multirow{1}{*}{\cite{Elstner1998}}         &\multirow{1}{*}{60}  &      N/A                 &    71.25  &   7.87   &  9.05   &     &   34.85   &   3.78   &   9.22   \\
  \addlinespace
  \hline
  \addlinespace
  \multirow{3}{*}{graphene nanoribbon}             &  \multirow{3}{*}{\cite{Elstner1998}}        &\multirow{3}{*}{52}  &       2                  &  3129.82  &  945.69  &  3.31   &     &  623.13   &  373.32  &   1.67   \\  
                                                   &                                             &                     &       3                  &  6781.11  & 1697.48  &  3.99   &     & 1438.13   &  567.75  &   2.53   \\
                                                   &                                             &                     &       4                  & 13602.81  & 2268.69  &  5.99   &     & 3118.92   &  748.01  &   4.17   \\
  \hline                                                                                                                                       
  \addlinespace
  \multirow{3}{*}{BN (18,18)-nanotube} & \multirow{3}{*}{\cite{Wahiduzzaman2013, Oliveira2015}}  &\multirow{3}{*}{72}  &       2                  &  5908.53  &  917.00  &  6.44   &     & 1774.76   &  478.60  &   3.71   \\
                                       &                                                         &                     &       3                  & 13505.67  & 2070.13  &  6.52   &     & 4358.30   &  719.79  &   6.05   \\
                                       &                                                         &                     &       4                  & 23206.95  & 2353.30  &  9.86   &     & 7383.50   &  968.84  &   7.62   \\
  \addlinespace                                                                                                                                                                       
  \hline
  \addlinespace
  \multirow{3}{*}{BeCl trilayer}       & \multirow{3}{*}{\cite{Wahiduzzaman2013, Oliveira2015}} & \multirow{3}{*}{12}  & $2 {\times} 2 $          &   1260.91 &   83.05  &  15.18  &     &  248.58   &  25.05   &   9.92   \\
                                       &                                                        &                      & $3 {\times} 3 $          &   5770.65 &  356.24  &  15.80  &     & 1207.91   &  56.14   &  21.52   \\
                                       &                                                        &                      & $4 {\times} 4 $          &  17546.44 &  612.03  &  28.67  &     & 3593.52   &  99.43   &  36.14   \\
  \addlinespace
  \hline
  \addlinespace
  \multirow{3}{*}{Si-doped graphene} & \multirow{3}{*}{\cite{Rauls1999}}                     & \multirow{3}{*}{32}  & $1 {\times} 1 $             &    446.98 &  72.51   &  6.16   &     &  102.65   &  26.10   &   3.93   \\ 
                                     &                                                       &                      & $2 {\times} 2 $             &   4461.40 & 285.72   & 15.61   &     &  850.49   & 105.45   &   8.06   \\ 
                                     &                                                       &                      & $3 {\times} 3 $             &  20877.22 & 863.71   & 24.17   &     & 4581.45   & 225.61   &  20.31   \\
  \addlinespace
  \hline
  \addlinespace
  \multirow{3}{*}{BN (zincblende)} & \multirow{3}{*}{\cite{Wahiduzzaman2013, Oliveira2015}}  & \multirow{3}{*}{2}   & $2 {\times} 2 {\times} 2$   &    34.26  &  20.08   &  1.71   &     &   12.68   &  15.77   &   0.80   \\  
                                   &                                                         &                      & $3 {\times} 3 {\times} 3$   &   180.83  &  80.09   &  2.26   &     &   51.45   &  54.93   &   0.94   \\  
                                   &                                                         &                      & $4 {\times} 4 {\times} 4$   &   929.98  & 188.96   &  4.92   &     &  222.13   & 129.90   &   1.71   \\
  \addlinespace
  \hline
  \addlinespace    
  \multirow{3}{*}{ZnO bulk}        & \multirow{3}{*}{\cite{Moreira2009}}                     & \multirow{3}{*}{4}   & $2 {\times} 2 {\times} 1$   &   106.34  &  13.62   &  7.81   &     &   31.62   &   8.32   &   3.80   \\  
                                   &                                                         &                      & $3 {\times} 3 {\times} 2$   &  1039.87  &  75.18   & 13.83   &     &  224.85   &  37.82   &   5.95   \\  
                                   &                                                         &                      & $4 {\times} 4 {\times} 3$   &  6041.28  & 225.84   & 26.75   &     & 2072.09   & 103.35   &  20.05   \\
  \addlinespace
  \hline
  \addlinespace                                        
  \multirow{3}{*}{4H SiC}          & \multirow{3}{*}{\cite{Wahiduzzaman2013, Oliveira2015}}  & \multirow{3}{*}{8}   & $2 {\times} 2 {\times} 1$   &  1116.68  &  67.73   & 16.49   &     &  275.73   &  35.69   &   7.73   \\  
                                   &                                                         &                      & $3 {\times} 3 {\times} 1$   &  4333.59  & 274.94   & 15.76   &     &  973.96   &  79.76   &  12.21   \\  
                                   &                                                         &                      & $4 {\times} 4 {\times} 1$   & 13570.27  & 548.14   & 24.76   &     & 2989.49   & 140.80   &  21.23   \\                                      
  \bottomrule
  \end{tabular}}
\end{table}
\begin{adjustwidth}{-1.5cm}{-1.5cm}
\begin{figure}[h]
  \centering
  \subfigure{\includegraphics[width=0.31\linewidth]{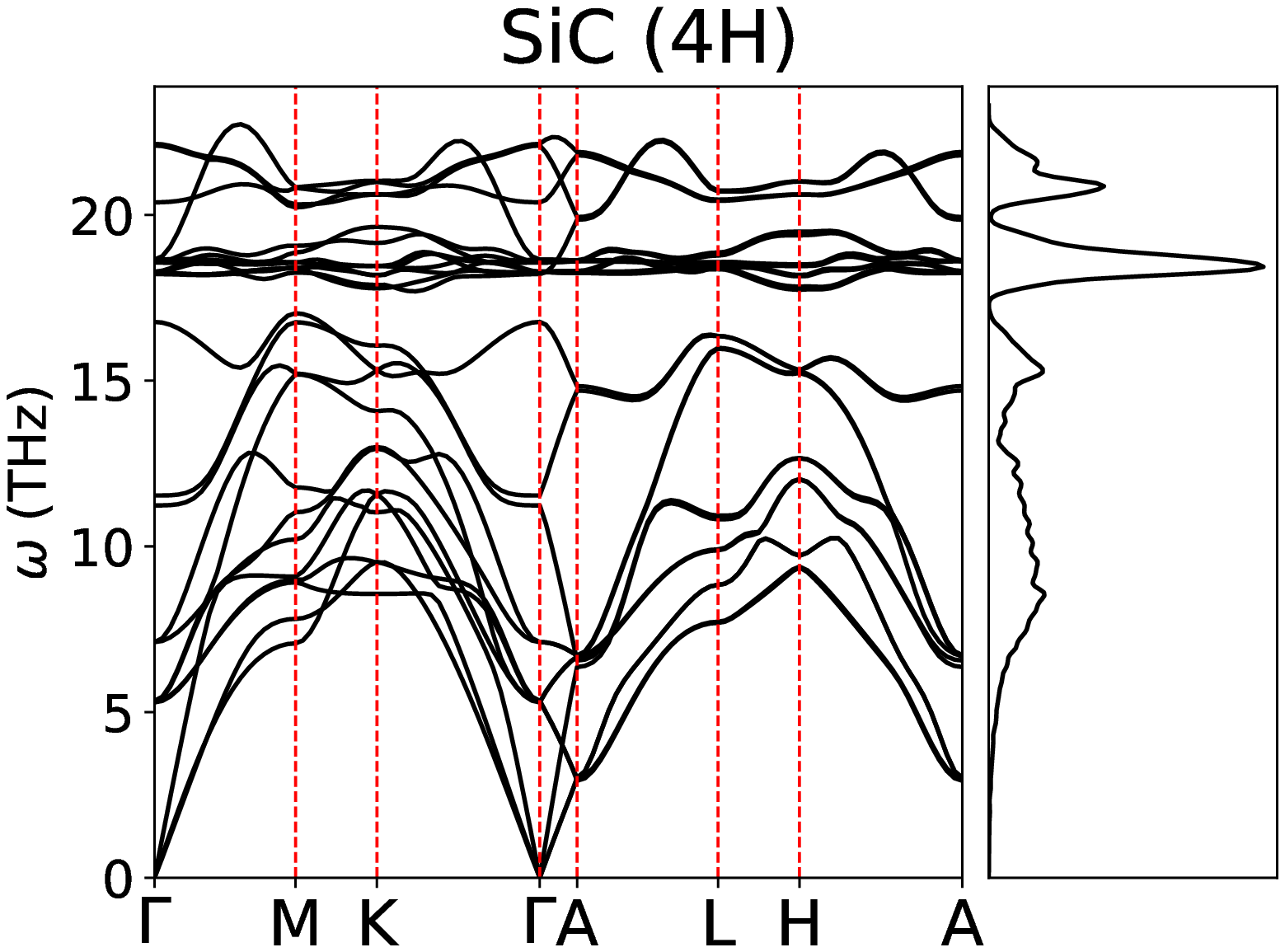}}
  \subfigure{\includegraphics[width=0.31\linewidth]{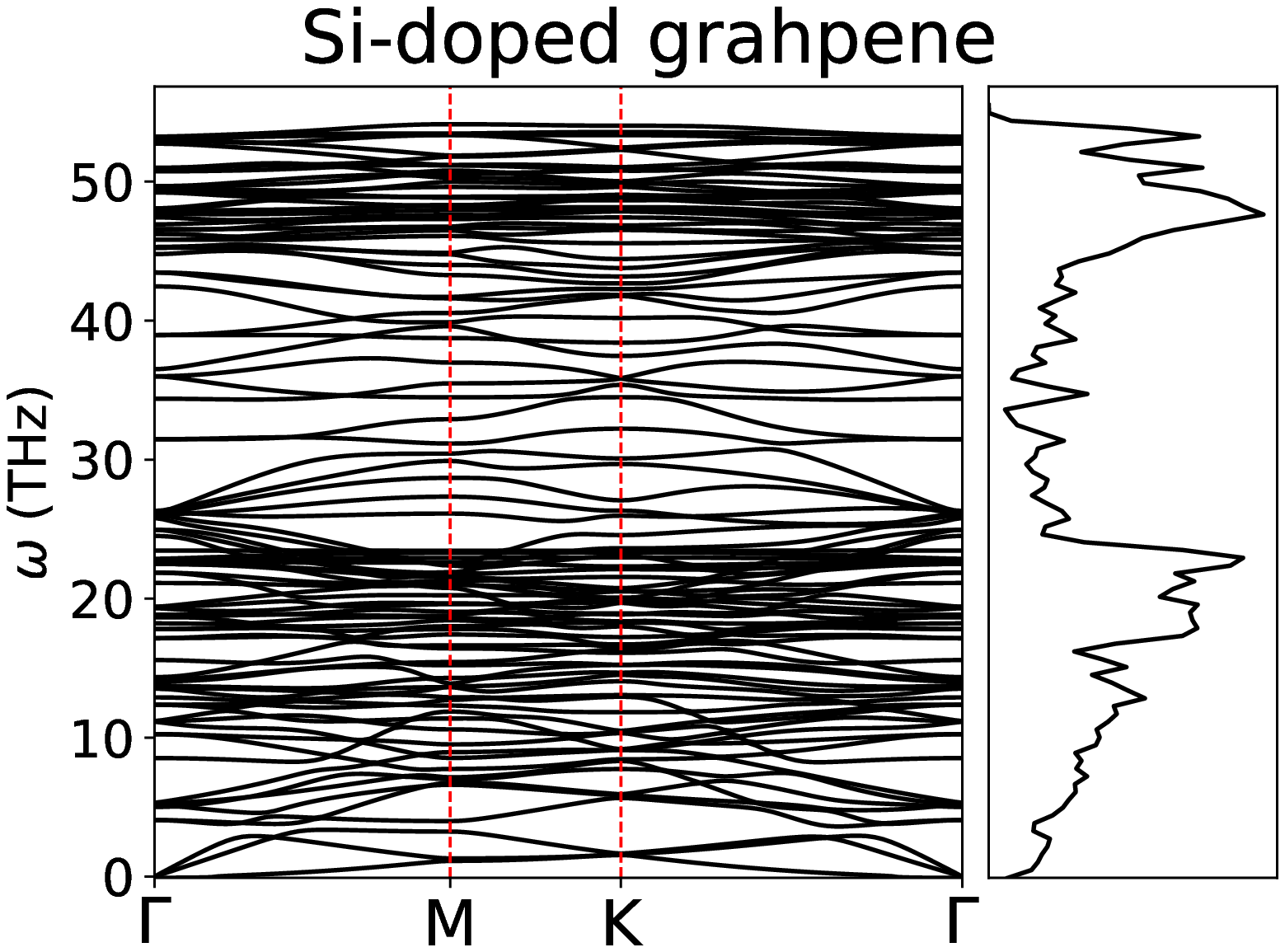}}
  \subfigure{\includegraphics[width=0.31\linewidth]{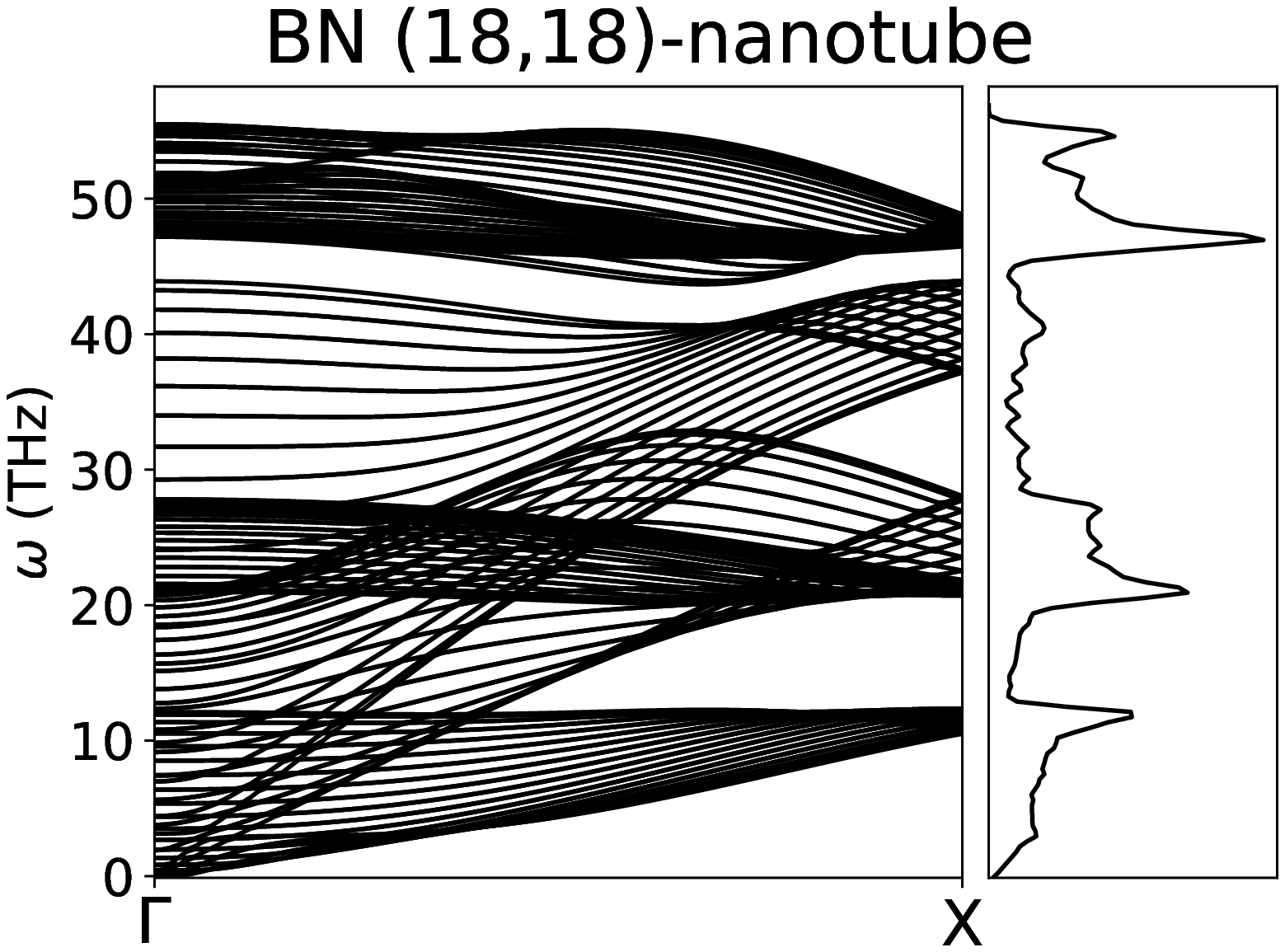}}
  \caption{Phonon dispersions and DOS of selected systems}\label{fig:phdisp}
\end{figure}
\end{adjustwidth}
\newpage
\section{Conclusions}\label{conclusion}
In summary, we have successfully derived the reciprocal-space approach of Hessian calculation within the SCC-DFTB framework.
This approach allows for the Hessian of periodic systems to be obtained accurately and without doing any calculations on supercells, 
while also providing some information about electron-phonon interactions, which can be of great importance in the study of transport phenomena \cite{PhysRevB.95.245210}.
The formulation presented in this paper effectively generalizes previous analytical methods of SCC-DFTB Hessian calculation \cite{Witek2004, Nishimoto2017},
which are less suited for periodic systems.
Its efficiency has been demonstrated by performing test calculations on various systems of all dimensions,
where it showed to be significantly faster than the numerical force-differentiation method, especially for large systems.
\\
\indent
We believe that further research on DFTB-based phonon calculation methods can open a gateway to an efficient ab-initio description of phonon-related properties
(such as electrical or thermal conductivity, Raman spectra, band-gap renormalization, superconductivity)\ of large systems, 
otherwise computationally too demanding for ordinary DFT methods, and perhaps even lead to more extensive DFTB parameters develpoment for solid-state applications.
This paper can be regarded as a first step in that direction.
In our future work, we plan various extensions of the theoretical formulation presented here (such as to DFTB3 framework and adding support for spin-polarization), 
as well as its application to problems of interest.
\section*{Acknowledgements}
We thank Pier Philipsen, Mirko Franchini, Robert R\"{u}ger, Augusto Oliveira, M. S. Ramzan and Stan van Gisbergen for useful discussions and technical support.
The financial support by the Deutsche Forschungsgemeinschaft (GRK 2247/1 (QM3)) is greatly appreciated.
\section*{Supporting Information}
The data that supports the findings of this study are available within the article [and its supplementary material].

\section*{Data Availability}
The data that support the findings of this study are available from the corresponding author upon reasonable request.
\begin{appendices}
\numberwithin{equation}{section}
\section{Derivatives of the DFTB matrix elements} \label{App A}
According to the Slater-Koster transformation rules \cite{Slater1954}, the two-center integrals $I_{ab}$,
between basis functions $a$ and $b$, located on atoms separated by vector $\vec{r}$, can be written as:
\begin{equation}
   I_{ab}^{} (\vec{r}) = \sum_{\tau} R_{ab}^{\tau} (r) A_{l(ab)}^{\tau} (\hat{\vec{r}})
\end{equation}
where $\tau$ is the index of the Slater-Koster integral,  $r \equiv \lVert \vec{r} \rVert$, and $\hat{\vec{r}} \equiv \vec{r} / r $. 
In the DFTB formalism, the radial functions $R_{ab}^{\tau} (r)$ are generally given on a numerical grid.
However, they can always be cast to an analytical form, for example, by spline interpolation.
Unlike $R_{ab}^{\tau} (r)$, the functions $A_{l(ab)}^{\tau} (\hat{\vec{r}}) $ depend only on the angular momenta of $a$ and $b$ basis functions and not on their radial shape.
Since they are given in a purely analytical form, their derivatives can be obtained easily.
For example, if $a$ is an $s$- and $b$ a $p$-type function, we have:
\begin{subequations}
  \begin{align}
  & A_{sp_i}^{\sigma} (\vec{\hat{r}} )  = \frac{r_i}{r} \\ 
  \partial_j & A_{sp_i}^{\sigma} (\vec{\hat{r}} ) = \frac{\delta_{ij}}{r} - \frac{r_i r_j} {r^2} \\
  \partial_k \partial_j & A_{sp_i}^{\sigma} (\vec{\hat{r}} ) = - \frac{1}{r^3} ( \delta_{ij} r_k + \delta_{ik} r_j +  \delta_{jk} r_i ) + 3 \frac{r_i r_j r_k} {r^5}
  \end{align}
\end{subequations}
where $r_i$ is the i-th component of $\vec{r}$ and $\partial_i \equiv \partial / \partial r_i$.
Similar expressions can be derived for all other combinations of angular momenta.
\\
Finally, the first and second derivative of $I_{ab}$ can be written as:
\begin{subequations} 
  \begin{align}
  \partial_i            I_{ab} (\vec{r}) = \sum_{\tau} \bigg[& \partial_r R_{ab}^{\tau} \frac{r_i}{r} \, A_{l(ab)}^{\tau} + 
                                                               R_{ab}^{\tau}  \, \partial_i A_{l(ab)}^{\tau}
                                                       \bigg]
       \\
  \partial_j \partial_i I_{ab} (\vec{r}) = \sum_{\tau} \bigg[& \left( \partial_r^{2} R_{ab}^{\tau} - \frac{\partial_r R_{ab}^{\tau}}{r} \right) \frac{r_i r_j}{r^2} \, A_{l(ab)}^{\tau} \,+\, 
                                                      R_{ab}^{\tau} \, \partial_i \partial_j A_{l(ab)}^{\tau} \nonumber
                                                      \\
                                                     & \quad +\, \frac{ \partial_r R_{ab}^{\tau} } {r} 
                                                     \Big(  \delta_{ij}  A_{l(ab)}^{\tau} +  r_i \partial_j A_{l(ab)}^{\tau} + 
                                                                                             r_j \partial_i A_{l(ab)}^{\tau}
                                                     \Big)
                                             \bigg]
  \end{align}
\end{subequations}
where the derivatives of $R_{ab}^{\tau}$ can be obtained analytically.
\section{Lattice summations of the DFTB $\gamma$-function} \label{App B}
The phase-modulated lattice sum of the $\gamma$ function is given by:
\begin{equation}\label{sum_def}
   \widetilde{\gamma}_{IJ}^{} (\vec{q}) \equiv \sum_{\vec{R}} e_{}^{i\vec{qR}} \gamma_{IJ}^{} (\vec{R} + \vec{u}_J^{} - \vec{u}_I^{})
\end{equation}
where $\vec{u}_X$ is the position vector of atom $X$, as defined in the original unit cell.\\
Adding and subtracting the Coulomb potential, the expression for $\gamma$-function can be written as:
\begin{equation}\label{gamma_split}
   \gamma_{IJ}^{} (\vec{r}) = \bigg( \gamma_{IJ}^{}(r) - \frac{1}{r} \bigg) + \frac{1}{r}
\end{equation}
Since $\gamma_{IJ}(r) \rightarrow 1/r$ as $ r \rightarrow \infty$, the first term in this expression is short-ranged, making the lattice summation over it straightforward.
This is not the case for the second term, so here we use the Ewald summation technique \cite{Ewald1921}, i.e., we split it into a short-ranged and a long-ranged part:
\begin{equation}
  \frac{1}{r} = \frac{ \text{erfc}(\alpha r) }{ r } + \frac{ \text{erf}(\alpha r) }{ r }
\end{equation}
% %
where $\alpha$ is an arbitrary positive real number.
The short-ranged term can be added to the first term on the RHS of \eqref{gamma_split}, whereas the (generalized) Poisson summation formula can be used for the long-ranged part.
The expression for $\widetilde{\gamma}(\vec{q})$ then becomes:
\begin{equation}
   \widetilde{\gamma}_{IJ}^{} (\vec{q}) = \sum_{\vec{R}} e_{}^{i\vec{qR}} \gamma_{IJ}^{\alpha} ( \vec{R} + \vec{u}_J^{} - \vec{u}_I^{} ) 
   \,+\, \sum_{\vec{G}} \widetilde{V}_{}^{\alpha} (\vec{G + q}; \vec{u}_{I}^{} - \vec{u}_J^{}) e_{}^{i(\vec{G+q})(\vec{u}_I - \vec{u}_J)}
\end{equation}
The second sum here runs over all reciprocal vectors $\vec{G}$, while $\alpha$ can be chosen to ensure good convergence of both sums.
$\gamma_{IJ}^{\alpha} (\vec{r})$ is a short-ranged function, given by:
\begin{equation}
   \gamma_{IJ}^{\alpha} (\vec{r}) \equiv  \gamma_{IJ}^{}(r) - \frac{ \text{erf}(\alpha r) }{ r } + \frac{2 \alpha}{\sqrt{\pi}} \delta_{r=0}^{}
\end{equation}
while the expression for $\widetilde{V}_{}^{\alpha} (\vec{k} ; \vec{r} )$ is more complicated, as it depends on the dimension of the underlying lattice
and on whether $\lVert \vec{k} \rVert$ is finite or not \cite{Martial2010, Porto2000}; see table \ref{V_alpha} for details.
\begin{table}[h]
   \renewcommand{\arraystretch}{2.8}
   \resizebox{1.0\textwidth}{!}{\begin{minipage}{\textwidth}
   \caption{ $\widetilde{V}_{}^{\alpha} (\vec{k} ; \vec{r} )$ for different dimensions. $\Omega$ is the measure of the underlying unit cell 
             (i.e., volume, area and length for three-, two- and one-dimensional systems, respectively).
             For the two-dimensional case, $z$ is the component of $\vec{r}$ perpendicular to the direction of the periodicity. 
             For the one-dimensional case, $\rho \equiv \sqrt{x^2 + y^2}$, where $x$ and $y$ are components of $\vec{r}$ perpendicular to the direction of the periodicity, 
             $\Gamma(u,v)$ is the upper incomplete gamma-function and $\gamma_{E}$ is the Euler-Mascheroni constant. \label{V_alpha}}
   \begin{tabular}{c  l  l}
     \toprule
     {dim}  &  \multicolumn{1}{c}{$\widetilde{V}_{}^{\alpha} (\vec{k} \neq \vec{0} ; \vec{r})$}  &   \multicolumn{1}{c}{$\widetilde{V}_{}^{\alpha} (\vec{k} = \vec{0} ; \vec{r})$} \\
     \hline
     \hline
      3     &  \( \displaystyle \frac{4 \pi}{\Omega} \frac{ e^{-k^2 / 4 \alpha^2} } { k^2 } \)   & \qquad 0 \\
               
      2     &  \( \displaystyle \frac{\pi}{\Omega} \frac{1}{k} 
                  \bigg[ e^{-kz} \text{erfc}\bigg( \frac{k} {2\alpha} - \alpha z \bigg) +
                         e^{ kz} \text{erfc}\bigg( \frac{k} {2\alpha} + \alpha z \bigg)
                  \bigg] \)                                                                     
               &
               \quad \; \( \displaystyle \frac{2\pi}{\Omega} \bigg[ z \, \text{erf}(\alpha z) + \frac{ e^{-\alpha^2 z^2} }{ \alpha \sqrt{\pi} } \bigg] \) \\
      1     &  \( \displaystyle \frac{1}{\Omega} 
                  \bigg[  \delta_{\rho \neq 0} \sum_{n=0}^{\infty} \frac{ (-1)^n } { 4^n n! } (k \rho)^{2n} \Gamma \Big(-n, \frac{k^2}{4 \alpha^2} \Big)
                        + \delta_{\rho =    0} \Gamma \Big(0, \frac{k^2}{4 \alpha^2} \Big)
                  \bigg] \)
               &
               \quad \( \displaystyle -\frac{1}{\Omega} \Big[ \gamma_{E} + \Gamma(0, \alpha^2 \rho^2) + \log(\alpha^2 \rho^2) \Big] \) \\
     \bottomrule
   \end{tabular}
   \end{minipage}}
\end{table}
\\
Lattice summations involving the derivatives of the $\gamma$-function (\eqref{d_gamma_lattsum} and \eqref{d2_gamma_E}) can be evaluated in a similar manner.
\end{appendices}
\newpage
%\bibliography{ref2.bib}
%\bibliographystyle{achemso}
%
\providecommand{\latin}[1]{#1}
\makeatletter
\providecommand{\doi}
  {\begingroup\let\do\@makeother\dospecials
  \catcode`\{=1 \catcode`\}=2 \doi@aux}
\providecommand{\doi@aux}[1]{\endgroup\texttt{#1}}
\makeatother
\providecommand*\mcitethebibliography{\thebibliography}
\csname @ifundefined\endcsname{endmcitethebibliography}
  {\let\endmcitethebibliography\endthebibliography}{}

\end{document}